\documentclass[10pt, aps, prl, amssymb, amsmath, amsfonts, showpacs, floatfix, twocolumn, twoside, a4paper, superscriptaddress, longbibliography, nofootinbib]{revtex4-2}

\usepackage{siunitx}
\usepackage{comment}
\usepackage{soul}
\usepackage[T3,T1]{fontenc}
\DeclareMathAlphabet{\mathpzc}{OT1}{pzc}{m}{it} % For \mathpzc fonts
\usepackage{mathrsfs} % For \mathscr fonts
\usepackage{bm} % For \bm fonts
\usepackage{microtype} % For nicer alignment of text
\usepackage{orcidlink}
\usepackage{multirow}
\usepackage{booktabs} % For better rules in the table
\usepackage{array} % To define the C column type
\usepackage{tabularx}
\usepackage{braket} % For \bra and \ket

\newcolumntype{C}{>{\centering\arraybackslash}X}

\usepackage{xcolor}
\usepackage{hyperref} % For handling hyperlinks and citation colors
\usepackage{nicefrac} % For \nicefrac command

\hypersetup{
	colorlinks=true,    % Colors the links instead of using boxes
	citecolor=blue,     % Sets the color of citations to blue
	linkcolor=red,      % Sets the color of internal links (e.g., sections) to red
	urlcolor=blue,     % Sets the color of URLs to green
}

\usepackage{graphicx} %
\graphicspath{{./Figures/}}
\usepackage{xspace}
\newcommand{\iu}{\mathrm{i}}
\newcommand{\dd}{\mathrm{d}}

\newcommand{\mb}{m_\phi}

\newcommand{\grappa}{Gravitation Astroparticle Physics Amsterdam (GRAPPA),\\ University of Amsterdam, 1098 XH Amsterdam, The Netherlands}

\newcommand{\aei}{Max Planck Institute for Gravitational Physics (Albert Einstein Institute),\\ Am Mühlenberg 1, D-14476 Potsdam, Germany}

\DeclareSymbolFont{mathtx}{OML}{txmi}{m}{it}
\DeclareMathAlphabet\mathbfcal{OMS}{cmsy}{b}{n}
\DeclareMathSymbol{v}{\mathalpha}{mathtx}{118}

\begin{document}
\title{Weighing dark matter particles with gravitational waves}

\author{Rodrigo Vicente}
\email{r.l.lourencovicente@uva.nl}	
\affiliation{\grappa}
\author{Francisco Duque}
\affiliation{\grappa}
\affiliation{\aei}
\author{Theophanes K. Karydas}
\affiliation{\grappa}
\author{Gianfranco Bertone}
\affiliation{\grappa}

\date{October 27, 2026}

\begin{abstract}
It is commonly assumed that the mass of dark matter particles cannot be determined from their gravitational effects when dark matter behaves as collisionless matter. Here we show that extreme-mass-ratio inspirals observable by millihertz gravitational-wave observatories such as the Laser Interferometer Space Antenna (LISA) could measure the mass of ultralight dark matter particles in sufficiently dense dark matter spikes. The separation of scales between the two black holes allows ultralight dark matter to form a collisionless density spike around the massive primary while retaining wave effects in its interaction with the stellar-mass companion. We incorporate wave-dependent accretion and dynamical friction into relativistic inspiral waveforms and jointly infer the dark matter particle mass, spike normalisation and binary parameters from simulated LISA signals. For massive black holes at the low-mass end of the LISA population ($M\sim  10^5 M_\odot$), our forecasts show that particle masses spanning $10^{-14}$--$10^{-11}\,\mathrm{eV}$ could be measured to within an order of magnitude.
\end{abstract}

\maketitle

%%%%%%%%%%%%%%%%%%%%%%%
%{\em Introduction.---}
%%%%%%%%%%%%%%%%%%%%%%
%
%
The microscopic nature of dark matter (DM) remains unknown despite compelling evidence for its existence from galaxy and cluster dynamics, gravitational lensing and cosmological observations~\cite{Bertone:2004pz, Clowe:2006eq, Bertone:2016nfn, Planck:2018vyg, Cirelli:2024ssz}. Searches for its non-gravitational interactions with ordinary matter have yet to identify its constituents. Purely gravitational probes offer a complementary approach~\cite{Bertone:2018krk}, but identifying DM through gravity requires observables sensitive to its microscopic properties, rather than to its distribution alone.
Extreme-mass-ratio inspirals (EMRIs) are sensitive probes of the matter surrounding massive black holes (MBHs). In these systems, a stellar-mass compact object spirals into a MBH of $10^5$--$10^7 \, M_\odot$, accumulating hundreds of thousands of orbital cycles in the millihertz frequency band accessible to the Laser Interferometer Space Antenna (LISA)~\cite{Babak:2017tow, Speri:2026ade, Pound:2021qin_selfforce, 1805.10385, LISA:2024hlh}. Accretion and gravitational scattering (dynamical friction) of surrounding matter perturb the orbit. Although weak compared to gravitational radiation reaction, these environmental interactions can accumulate a measurable phase shift over years of observation.
Previous studies have shown that EMRI observations can distinguish between environments, including collisionless DM spikes, boson clouds and accretion disks, and reconstruct their macroscopic properties~\cite{Cole:2022yzw, Hannuksela:2018izj, Hannuksela:2019vip, Kim:2022mdj}. Distinguishing the constituents of environments with the same distribution is more challenging. Different DM candidates, from elementary particles to primordial black holes, can behave as collisionless matter around a MBH and produce similar phase-space distributions. Reconstructing such a distribution would therefore reveal the environment without necessarily identifying its microscopic constituents.

\begin{figure}
    \centering
    \includegraphics[width=\linewidth]{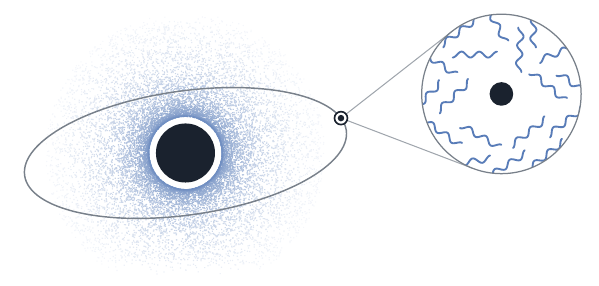}
    \caption{
    \textbf{Probing the wave nature of dark matter with extreme-mass-ratio inspirals.}
    A stellar-mass black hole spirals into a much more massive black hole surrounded by a dark matter spike. The large difference between the two black hole masses allows dark matter to behave as collisionless particles around the primary while retaining its wave nature in interactions with the smaller companion. The resulting wave effects modify the accretion and gravitational drag experienced by the companion, leaving an imprint on the gravitational-wave signal that depends on the dark matter particle mass.}
    \label{fig:scale-hierarchy}
\end{figure}

For ultralight bosonic DM~\cite{Hui:2016ltb, Hui:2021tkt, Ferreira:2020fam}, the separation between the two black hole scales in an EMRI provides access to the particle mass even when the surrounding distribution is collisionless. The gravitational scales of the primary and companion are $R_M\equiv GM/c^2$ and $R_m\equiv Gm/c^2$, with $R_m/R_M\lesssim10^{-4}$. A boson whose reduced Compton wavelength, $\lambdabar_C\equiv\hbar/(m_\phi c)$, satisfies
\begin{equation}\label{eq:hierarchy}
    R_m\lesssim\lambdabar_C\ll R_M\,,
\end{equation}
can behave as collisionless matter around the primary while retaining wave effects in its interaction with the companion (Fig.~\ref{fig:scale-hierarchy}). For typical LISA EMRIs, this hierarchy corresponds approximately to particle masses spanning six orders of magnitude
\begin{equation}\label{eq:mass-window}
10^{-16}\,\Big(\frac{10^6M_\odot}{M}\Big)\ll m_\phi/\mathrm{eV}\lesssim10^{-11}\,\Big(\frac{10 M_\odot}{m}\Big)\,.
\end{equation}
On the scale of the primary, the adiabatic growth of a MBH can compress a pre-existing collisionless halo into a steep density spike~\cite{Gondolo:1999ef}. Its relativistic phase-space distribution follows from conserving orbital actions during this growth~\cite{Sadeghian:2013laa, Ferrer:2017xwm}. Within this formation scenario, we find that the inner density and torque profiles have approximately universal radial shapes over the radii sampled by LISA EMRIs, allowing their dependence on the initial halo to be captured primarily by a single spike normalisation. On the scale of the companion, wave effects make the accretion and scattering cross-sections depend on the DM particle mass~\cite{Vicente:2022ivh, Traykova:2023qyv, Traykova:2021dua}. The spike normalisation sets the overall strength of the torque, whereas the particle mass also changes its radial dependence, allowing the two properties to be constrained jointly.
Here, we incorporate this environmental torque into relativistic inspiral waveforms and use Bayesian inference on simulated LISA observations to constrain the dark matter particle mass and spike normalisation jointly with the binary parameters. We model DM as a free, minimally coupled scalar and consider quasi-circular EMRIs into non-spinning MBHs.
For low-mass primaries embedded in dense spikes, we identify the particle masses in the range \(10^{-16}\)–\(10^{-11}\,\mathrm{eV}\) that can be measured within one order of magnitude, and establish how this reach depends on the binary parameters, initial host halo (where the MBH grows) and observation duration. Outside the regime of mass sensitivity, the signal can retain information about the spike even when the particle mass is only bounded or remains unconstrained.

\section{Results}
\subsection{The inner spike profiles have approximately universal shapes}

In principle, the spike depends on the initial halo profile, and that dependence would have to be carried through to the waveform. We find that it largely does not (Methods). Within the adiabatic growth scenario, the inner spike density and environmental torque profiles have nearly the same radial shape, over the radii probed by our inspirals, for every initial cusp we tested. What changes is only an overall scale, which we absorb into a spike normalisation \(\bar \rho\), leaving one astrophysical parameter to be inferred together with the DM particle mass.

We define the reference spike around a MBH of mass \(M_0=10^6\,M_\odot\), grown within an initial cusp of slope~\(\gamma_0=1\) and density \(150\,\mathrm{GeV}/c^2\,\mathrm{cm}^{-3}\) at \(100\,\mathrm{pc}\) (which matches the inner density of the Milky-Way-like Hernquist halo adopted in previous relativistic spike calculations~\cite{Sadeghian:2013laa, Ferrer:2017xwm}). Its normalisation is \(\bar\rho=1\) by construction, and its density peaks at $2.4\times10^{19}\,\mathrm{GeV}/c^2\,\mathrm{cm}^{-3}$ at $r\simeq7.7R_M$. Writing its density profile as \(\rho_0(r)\), the density of another spike is $\rho(r)/\rho_M \simeq \bar \rho\, \rho_0[(M_0/M) r]/\rho_{M_0}$, where $\rho_M\equiv M/R_M^3$. The torque admits an analogous rescaling (Methods).  

For a MBH grown within a self-gravitating initial cusp, $\rho_i(r)\simeq\rho_{100\,{\rm pc}}(r/100\,\mathrm{pc})^{-\gamma}$, the spike normalisation $\bar \rho$ is related to $\rho_{100\, \rm pc}$ and $\gamma$ through a simple analytic expression (Eq.\eqref{eq:DensityNorm}, Methods). The re-scaling of the environmental torque through this single factor reproduces the exact result to better than \(5\%\) within $r\leq 30 R_M$, for every slope \(0.25\leq\gamma\leq1.5\), an accuracy error much smaller than our forecasts require (as we will see in the next section).

Stellar-kinematic studies of cuspy dwarf galaxies infer DM densities of order $10\,\mathrm{GeV}/c^2\,\mathrm{cm}^{-3}$ at $150\,\mathrm{pc}$~\cite{Read:2018pft}, while observationally constrained Milky Way halo models yield densities of order $10^2\,\mathrm{GeV}/c^2\,\mathrm{cm}^{-3}$ when evaluated at $100\,\mathrm{pc}$~\cite{McMillan:2017}. Cosmological hydrodynamical simulations show that baryonic contraction and feedback can substantially modify the inner density and slope, depending on the host's formation history~\cite{Calore:2015oya, Maccio:2020svl, Gnedin:2004cx}. Motivated by these estimates, we explore cuspy halos with
\begin{equation}
    \rho_{100\,{\rm pc}}\sim10\,\text{--}\,10^3\,\mathrm{GeV}/c^2\,\mathrm{cm}^{-3}\,,
    \label{eq:plausible-range}
\end{equation}
taking \(\gamma=1\) as an NFW-like reference and \(\gamma=1.5\) as an optimistic steep cusp. These choices define the host halos used below. Their conversion into spike normalisations assumes adiabatic MBH growth, while their applicability to EMRI hosts also depends on the assembly history and subsequent survival of the spike (Discussion).

\subsection{A low-mass LISA EMRI constrains both the DM particle mass and the spike density}

We jointly infer the DM particle mass, spike density and binary parameters from simulated LISA observations of quasi-circular inspirals into non-spinning MBHs. We incorporate the environmental torque into relativistic waveforms generated with FastEMRIWaveforms~\cite{Chua:2020stf,Katz:2021yft,Speri:2023jte,Chapman-Bird:2025xtd} and sample the posterior using the Markov-chain Monte Carlo sampler \(\texttt{eryn}\)~\cite{Karnesis:2023ras}. The simulations do not contain detector-noise, while the likelihood includes LISA’s noise power spectral density (see Methods for more details on the waveform model and the data analysis setup).

\begin{figure*}
    \centering
    \includegraphics[width=\linewidth]{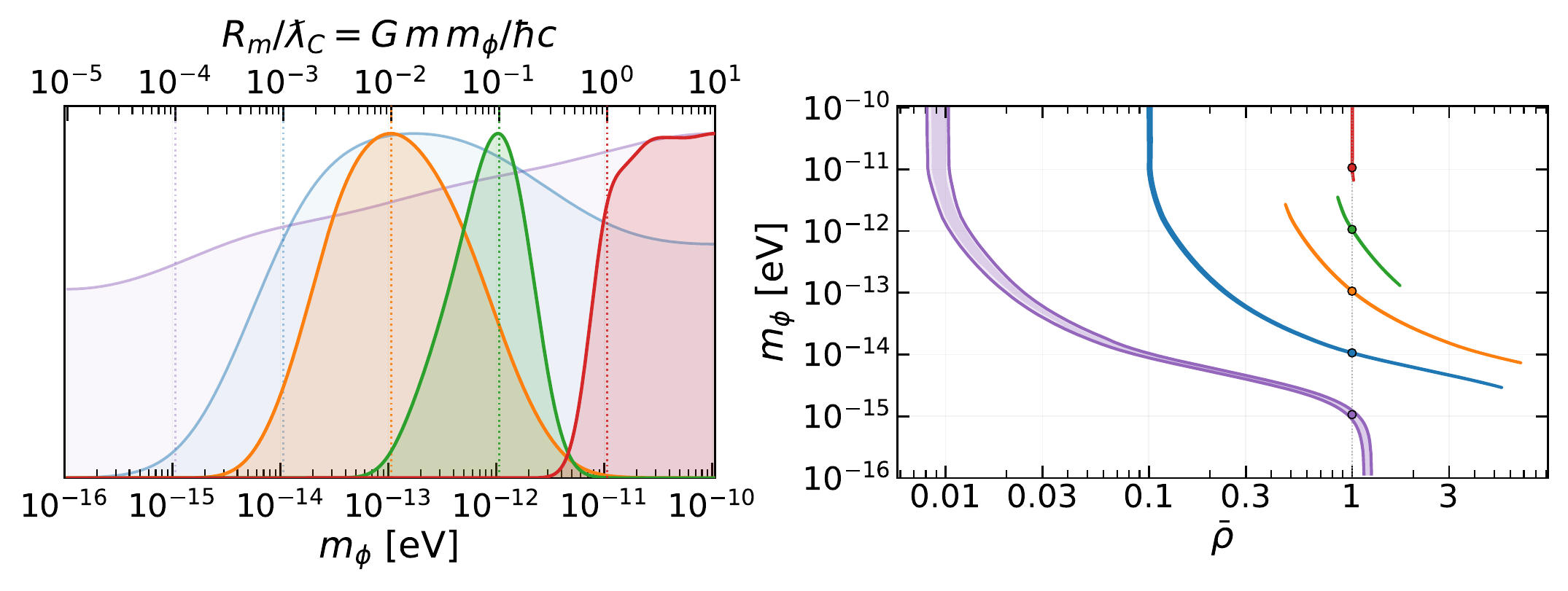}
    \caption{\textbf{Constraints on the DM particle mass and spike normalisation for a low-mass EMRI in a spike.} \textit{Left}: Marginalized posteriors of $\mb$ for five injected particle masses (marked by dotted lines). \textit{Right}: Joint $90\%$ credible regions in the $(\bar\rho,\mb)$ plane, with dots marking the injected values.
    The fiducial EMRI has detector-frame masses $M=2\times10^{5}\,M_\odot$ and $m=18\,M_\odot$, a non-spinning MBH and a quasi-circular orbit. We consider the final three years before plunge assuming a signal-to-noise ratio of $50$, corresponding to $d_L=2.47\,\mathrm{Gpc}$ and $z=0.43$ (source-frame masses $1.40\times10^5\,M_\odot$ and $12.6\,M_\odot$). The spike has $\bar\rho=1$, corresponding to $\rho_{100\,{\rm pc}}\simeq1.1\times10^3\,\mathrm{GeV}/c^2\,\mathrm{cm}^{-3}$ for $\gamma=1$, or $5.6\,\mathrm{GeV}/c^2\,\mathrm{cm}^{-3}$ for $\gamma=1.5$ [Eq.~\eqref{eq:DensityNorm}]. %These densities are approximately $7$ and $0.04$ times the reference value, respectively.
    }
    \label{fig:mass-posteriors}
\end{figure*}

We consider the final three years before plunge of a low-mass EMRI with detector-frame masses $M=2\times10^5\,M_\odot$ and $m=18\,M_\odot$, embedded in a spike with $\bar\rho=1$. We assume a signal-to-noise ratio (SNR) of $50$, which places the source at a luminosity distance $d_L=2.47\,\mathrm{Gpc}$ (redshift $z=0.43$). Across injections, we vary only the DM particle mass. The observation begins at orbital radius $r\simeq26R_M$ and spans approximately $1.7\times10^5$ orbital cycles. 
This relatively light primary is a favourable source because it allows the observation to probe larger orbital separations (in units of $R_M$), where the environmental torque is stronger relative to gravitational radiation reaction (Fig.~\ref{fig:torque-ratio}, Methods). Depending on the particle mass, the spike advances the orbital phase by approximately $30$--$2000$ cycles relative to the vacuum inspiral. The binary masses and distance are representative of low-mass systems in the EMRI catalogue of Ref.~\cite{Babak:2017tow}.

Figure~\ref{fig:mass-posteriors} shows the joint constraints on the DM particle mass and spike normalisation for that low-mass EMRI.
For dark matter particle masses of $1.1\times10^{-13}\,\mathrm{eV}$ and  \linebreak $1.1\times10^{-12}\,\mathrm{eV}$, the marginalized mass posteriors peak near the true values and give two-sided constraints, recovering the mass to within about an order of magnitude. Across this window the environmental torque profile varies with the particle mass by more than the \(5\%\) accuracy of our spike rescaling (Fig.~\ref{fig:torque-ratio}, Methods). 
%Their respective $90\%$ credible intervals are $1.7\times10^{-14}$--$1.3\times10^{-12}\,\mathrm{eV}$ and $2.0\times10^{-13}$--$2.5\times10^{-12}\,\mathrm{eV}$. 
The particle mass and spike normalisation are strongly correlated, broadening their marginal constraints. Increasing $\bar\rho$ rescales the torque, whereas changing $m_\phi$ also changes its radial dependence (Fig.~\ref{fig:torque-ratio}, Methods). This partially breaks the degeneracy, although shifts in the binary parameters can absorb part of the associated phase evolution (Fig.~\ref{fig:full-posterior}, Methods). 
%Around these masses, the degeneracy weakens towards larger $m_\phi$, tightening the constraints on both DM parameters.

Towards lighter particles the torque flattens onto its long-wavelength plateau, some 70 times weaker than in the collisionless limit (Fig.~\ref{fig:torque-ratio}, Methods), and stops carrying information about the mass. Already at \(1.1\times10^{-14}\,\mathrm{eV}\) the posterior runs towards heavier particles in less dense spikes, leaving only a lower bound, and at \(1.1\times10^{-15}\,\mathrm{eV}\) the data no longer constrain the mass at all. The effect of the spike itself remains detectable. The lightest particles we simulate also lie at the edge of our collisionless treatment of the spike, and we include them only to delimit where sensitivity ends (Methods). Forecasts at lower particle masses require including wave effects in the distribution around the primary~\cite{Kim:2022mdj}. Within a narrow mass range, $4\times 10^{-16}\lesssim\mb/\mathrm{eV}\lesssim 10^{-15}$, scalar particles could also be probed through superradiance~\cite{Brito:2015oca}, provided the MBH has undergone a phase of rapid rotation.

Towards heavier particles, the interaction approaches the collisionless limit and the cross-sections lose their dependence on the particle mass. Increasing the mass from \(1.1\times10^{-11}\) to \(1.1\times10^{-10}\,\mathrm{eV}\) changes the accumulated dephasing relative to vacuum by less than \(1\%\), compared with an increase by a factor of four between \(1.1\times10^{-14}\) and \(1.1\times10^{-13}\,\mathrm{eV}\). At an injected mass of \(1.1\times10^{-11}\,\mathrm{eV}\), the posterior is consequently flat towards higher masses, yielding only a lower bound. Despite this loss of sensitivity to the particle mass, the spike produces approximately \(2000\) cycles of dephasing relative to vacuum, and its density is constrained with a precision better than \(1\%\).

\begin{figure*}
    \centering
    \includegraphics[width=\linewidth]{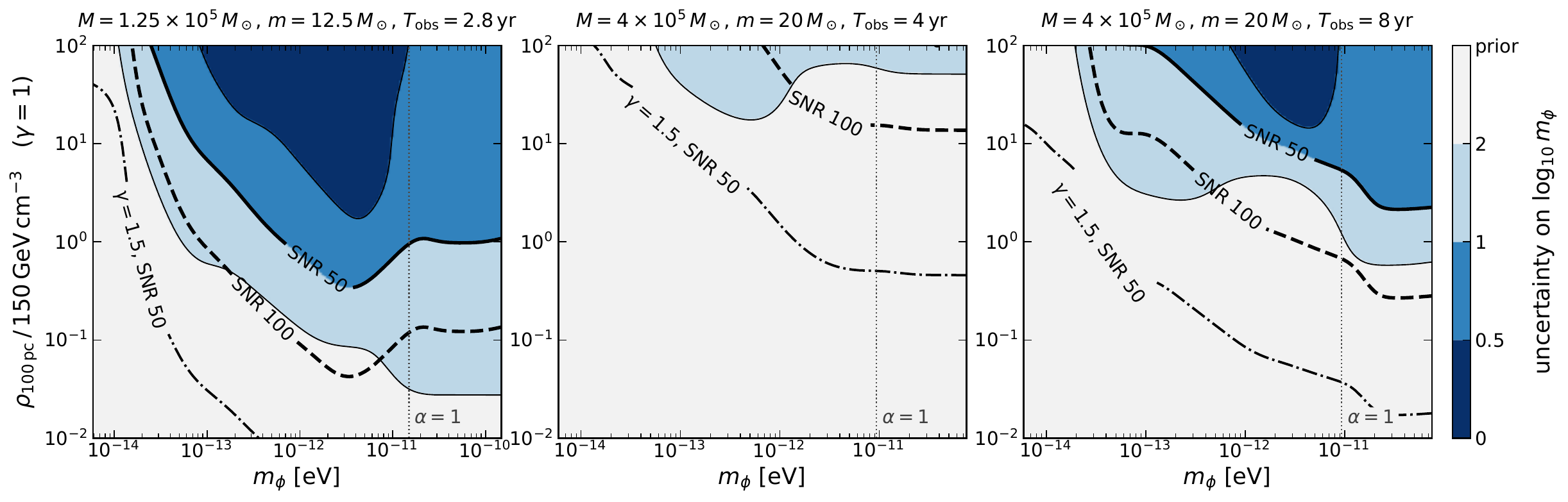}
    \caption{\textbf{Initial halo densities required for a measurement of the DM particle mass.}
    Colour shows the estimated width of the $90\%$ credible interval of $\log_{10}m_\phi$, for an initial cusp slope $\gamma=1$ and $\mathrm{SNR}=50$. The axes give the DM particle mass and initial halo density at $100\,\mathrm{pc}$, in units of the reference value $150\,\mathrm{GeV}/c^2\,\mathrm{cm}^{-3}$.
    \textit{Left}: EMRI with detector-frame masses $M=1.25\times10^{5}\,M_\odot$ and $m=12.5\,M_\odot$, observed for $2.8$ years. \textit{Middle and right}: EMRI with detector-frame masses $M=4\times10^{5}\,M_\odot$ and $m=20\,M_\odot$, observed for $4$ and $8$ years, respectively. Lines mark the threshold for a credible interval spanning one order of magnitude: $\gamma=1$ at $\mathrm{SNR}=50$ (solid) and $100$ (dashed), and $\gamma=1.5$ at $\mathrm{SNR}=50$ (dash-dotted).}
    \label{fig:measurable-band}
\end{figure*}

These results illustrate the possible constraints on the particle mass and spike normalisation for a favourable low-mass EMRI. The chosen spike normalisation requires either a high density halo, \(\rho_{100\,\mathrm{pc}}\simeq1.1\times10^3\,\mathrm{GeV}/c^2\,\mathrm{cm}^{-3}\) for \(\gamma=1\), or a steep cusp, with \(\rho_{100\,\mathrm{pc}}\simeq5.6\,\mathrm{GeV}/c^2\,\mathrm{cm}^{-3}\) for \(\gamma=1.5\). We next examine how the prospects depend on the binary masses, host halo properties and observation duration.

\subsection{How the sensitivity to the particle mass depends on the binary, host halo, and observation duration}

To explore the parameter space efficiently, we estimate the mass constraints from the residual orbital phase alone, refitting the spike normalisation and the binary parameters at each trial particle mass (Methods). The resulting interval widths agree with the full MCMC analysis to within \(30\%\) in every case we tested.

Figure~\ref{fig:measurable-band} shows how dense the initial host halo must be for the \(90\%\) credible interval on the particle mass to narrow to a single order of magnitude. We consider two binary systems, one lighter and one heavier than the fiducial system studied before, and compare four- and eight-year observations for the heavier binary. The longer observation lies within a possible extension of LISA operations to ten years~\cite{LISA:2024hlh}. We express the halo dependence through the density at $100\,\mathrm{pc}$, using $150\,\mathrm{GeV}/c^2\,\mathrm{cm}^{-3}$ as the reference value, and the initial cusp slope $\gamma$.

For the lighter binary, \(M=1.25\times10^5\,M_\odot\) and \(m=12.5\,M_\odot\), we consider the final 2.8 years before plunge at \(\mathrm{SNR}=50\). For \(\gamma=1\), at the reference density we can have an order-of-magnitude measurement on the particle mass from approximately \(7\times10^{-13}\,\mathrm{eV}\) up to the start of the collisionless limit. A steeper initial cusp, \(\gamma=1.5\), extends the lower edge of this range to \(2\times10^{-14}\,\mathrm{eV}\).

The heavier binary, \(M=4\times10^5\,M_\odot\) and \(m=20\,M_\odot\), requires substantially denser host halos over a four-year observation at the same SNR. For \(\gamma=1\), a particle mass measurement requires more than a hundred times the reference density, above our adopted plausible range~\eqref{eq:plausible-range}. For \(\gamma=1.5\), measurements extend down to about~\(1.5\times10^{-12}\,\mathrm{eV}\) at the reference density. Within that range, a four-year observation of this heavier EMRI therefore depends on a steep initial cusp. 

Higher SNR lowers the density needed for a mass measurement, and steeply. Within our approximation the constraint depends on \(\bar\rho\) and the SNR only through their product, so the minimum spike normalisation scales as \(\mathrm{SNR}^{-1}\), and since \(\rho_{100\,\mathrm{pc}}\propto\bar\rho^{4-\gamma}\) at fixed binary parameters and cusp slope, the halo density threshold scales as \(\mathrm{SNR}^{-(4-\gamma)}\). Doubling the SNR therefore lowers it by a factor of eight for \(\gamma=1\). For the lighter binary at the reference density, raising the SNR from \(50\) to \(100\) carries the lower edge of the measurable range from about \(7\times10^{-13}\) to \(9\times10^{-14}\,\mathrm{eV}\).

Extending the heavier binary's observation from four to eight years also substantially improves the prospects. For \(\gamma=1\) at \(\mathrm{SNR}=50\), the minimum required density falls from more than a hundred to six times the reference value. The corresponding density, \(\rho_{100\,\mathrm{pc}}\simeq900\,\mathrm{GeV}/c^2\,\mathrm{cm}^{-3}\), lies near the upper end of our adopted plausible range. For \(\gamma=1.5\), measurements extend down to about \(3.5\times10^{-14}\,\mathrm{eV}\) at the reference density.

This improvement comes primarily from the longer span of orbital evolution. The additional four years probe orbital separations of \(21\)–\(24~R_M\), where the environmental torque is stronger relative to gravitational radiation reaction, and triple the accumulated dephasing relative to vacuum. They raise the total SNR by only about \(5\%\), because the signal is weaker there and lies at lower frequencies, where LISA is less sensitive. What they add instead is phase evolution at separations the late inspiral does not cover, which shifts in the binary parameters cannot absorb. Longer observations can therefore bring the density required for a particle-mass measurement into the plausible range even when they add little SNR.

To sum up, our results show that low-mass EMRIs offer favourable prospects for measuring the mass of dark matter particles across a range of host halo properties, while heavier EMRIs require either a steeper initial cusp or a longer observation time.

%%%%%%%%%%%%%%%%%%%%%%%
\section{Discussion}
%%%%%%%%%%%%%%%%%%%%%%%
%
We have presented a gravitational-wave probe of DM microphysics using EMRIs in galactic nuclei. Our forecasts show that LISA observations of low-mass EMRIs could measure the mass of an ultralight scalar DM particle if the binaries are embedded in sufficiently dense spikes. Such spikes form through the adiabatic growth of MBHs within cuspy halos. The sensitivity to the particle mass arises from wave effects in gravitational scattering and accretion by the stellar-mass companion, which may coexist with a collisionless behavior around the MBH. These observations could constrain a microscopic property of DM jointly with its macroscopic distribution around MBHs on scales of $\lesssim 10^{-6}\,\mathrm{pc}$ (much smaller than galaxy-scale dynamical probes, which are sensitive to $\gtrsim 0.1\,\mathrm{kpc}$ scales).

The spike normalisation (which controls the strength of the effect on the EMRI) depends on the host halo and the epoch of substantial MBH growth. Extrapolating NFW profiles to $100\, \mathrm{pc}$, using the concentration–mass–redshift relation of Ref.~\cite{Dutton:2014xda} gives \(\rho_{100\,\mathrm{pc}}\simeq75\,\mathrm{GeV}/c^2\,\mathrm{cm}^{-3}\) for a \(10^{11}M_\odot\) halo at \(z=5\). A larger \(10^{12}M_\odot\) halo at the same epoch gives \(\rho_{100\,\mathrm{pc}}\simeq180\,\mathrm{GeV}/c^2\,\mathrm{cm}^{-3}\), rising to \(330\,\mathrm{GeV}/c^2\,\mathrm{cm}^{-3}\) for a concentration \(2\sigma\) above the median~\cite{Dutton:2014xda}. These examples illustrate how halo properties and growth epoch populate much of the density range considered here [cf.~\eqref{eq:plausible-range}]. Baryonic contraction can provide further enhancement towards its upper end~\cite{Calore:2015oya, Maccio:2020svl}.

Baryonic contraction can steepen the DM profile, depending on the central baryonic distribution~\cite{Gnedin:2004cx,Calore:2015oya,Schaller:2015mua}. This motivates exploring slopes beyond the NFW-like reference, with \(\gamma=1.5\) serving as an optimistic benchmark. Our forecasts show that such steep cusps substantially reduce the density needed for a measurement of the particle mass. Connecting this benchmark to specific contracted halo scenarios would require extending~\eqref{eq:plausible-range}, including the baryonic potential when it dominates the initial orbits supplying the spike.

The survival of a spike depends on the cumulative effect of previous inspirals. For a Sgr A*-like primary, \citet{Karydas:2026ctq} find that 270 mergers with \(10\,M_\odot\) companions reduce the inner density by 18\%, while the corresponding calculation of \citet{Sharpe:2026nqq} gives a reduction of 50\%. Larger depletion is predicted for lighter MBHs~\cite{Karydas:2026ctq,Sharpe:2026nqq}. The outcome depends on the number and orbital distribution of past EMRIs, with quasi-circular inspirals producing substantially stronger depletion. Ref.~\cite{Babak:2017tow} adopts a model leading to an inspiral rate of $\mathcal{O}(10)\,\mathrm{Gyr^{-1}}$ around MBHs of mass \(10^5\)–\(10^6M_\odot\). Such low rates would have a small impact on spikes around MBHs of mass~$\gtrsim 3\times10^5 M_\odot$~\cite{Karydas:2026ctq,Sharpe:2026nqq}. Nevertheless, previous EMRIs should be considered when connecting the density requirements obtained in our forecasts (Fig.~\ref{fig:measurable-band}) to a population of EMRI hosts.

Major mergers between comparable-mass MBHs cause more severe damage to spikes~\cite{Merritt:2002vj}. However, subsequent gas accretion could increase the central density again. The Sołtan argument indicates that accretion supplied most of the mass in the MBH population~\cite{Soltan:1982vf}, while demographic studies find substantial growth at \(z\lesssim2\), particularly for lower-mass MBHs~\cite{Shankar:2004ir, Marconi:2003tg}. Such growth, if it follows the last disruptive merger, provides an opportunity for spike regrowth. \citet{Caiozzo:2025mye} show that sufficiently large adiabatic growth can recover the spike distribution function, with the required growth depending on the radial extent of the depletion. The recovered spike could then survive in a host with a quiet subsequent merger history. Such quiet evolutions are supported by cosmological simulations, which show that approximately \(47\)–\(60\%\) of present-day \(10^{11}\)–\(10^{12}M_\odot\) haloes do not experience a major halo merger (with mass-ratio larger than 0.3) from \(z=2\)~\cite{Fakhouri:2010}.

Energy exchange via dynamical friction and mass accretion onto the companion can modify the spike distribution function and reduce the accumulated dephasing relative to the static spike approximation~\cite{Kavanagh:2020cfn, Kavanagh:2024lgq, Karydas:2024fcn}. Our analysis neglects such effects, as they are expected to be weaker for smaller mass ratios. While our forecasts show that the sensitivity to the particle mass improves for lighter MBHs, this trend may change when the feedback on the spike becomes relevant. Extending our results to intermediate mass black hole primaries would require a consistent treatment of such effects.

Our forecasts focus on quasi-circular inspirals around non-spinning MBHs. Extending the analysis to eccentric inspirals around spinning primaries would capture richer orbital evolution and additional harmonics. These features may help distinguish environmental effects from changes in binary parameters and improve LISA's constraints on the DM particle mass.

Although our forecasts concern free, minimally coupled scalars, the scattering framework of Ref.~\cite{Vicente:2022ivh} can be extended to other ultralight bosons, including vectors and tensors. The dependence of wave absorption and scattering on the boson spin could provide additional information about its microscopic identity. Establishing whether these differences remain distinguishable after jointly fitting the particle mass, spike density and binary parameters requires a dedicated analysis. We leave this and other aspects discussed above for future work.

%%%%%%%%%%%%%%%%%%%%%%%
\section{Methods}
%%%%%%%%%%%%%%%%%%%%%%%
%
Hereafter we use units with $G=c=1$, and a mostly positive metric signature.

%%%%%%%%%%%%%%%%%%%%%%%
\subsection{Waves behaving as particles}
%%%%%%%%%%%%%%%%%%%%%%%
%
Ultralight bosonic DM typically has large occupation numbers in the phase-space regions relevant to galactic halos, allowing a classical field description~\cite{Hui:2016ltb, Hui:2021tkt, Ferreira:2020fam}. Nevertheless, when its wavelength is sufficiently short, its macroscopic evolution admits an effective collisionless particle description. 
%Here we review how such effective description arises and discuss it applicability to dark matter spikes made of ultralight bosons.

We consider a free, minimally coupled real scalar field $\phi$ of mass $\mb$, obeying
\begin{equation}
    \left[\Box_g-(\mb/\hbar)^2\right]\phi=0,
    \qquad
    \Box_g\equiv g^{\mu\nu}\nabla_\mu\nabla_\nu.
    \label{eq:kg-wave-particle}
\end{equation}
Locally, a WKB component has the form $\phi=\operatorname{Re}[A\,e^{\iu S/\hbar}]$, with slowly varying real amplitude $A$ and four-momentum $p_\mu\equiv\nabla_\mu S$. The leading orders of the short-wavelength expansion give
\begin{equation}
    p^\mu p_\mu=-\mb^2, \qquad \nabla_\mu(A^2p^\mu)=0.
\end{equation}
Since $p_\mu$ is a gradient, the mass-shell relation implies $p^\mu\nabla_\mu p_\nu=\tfrac12\nabla_\nu(p^\mu p_\mu)=0$. So, the rays follow timelike geodesics, while the amplitude equation describes conservation of their weight. Equivalently, the phase-averaged stress-energy tensor is $\langle T^{\mu\nu}\rangle\simeq A^2p^\mu p^\nu/(2\hbar^2)$, so each component behaves as a pressureless stream.

A virialised halo or spike requires a superposition of components with different momenta. After suitable phase-space smoothing, their distribution $F(x^\mu,p^i)$ obeys the collisionless Boltzmann equation at leading order~\cite{Widrow:1993qq, Widrow:1996eq},
\begin{equation}
    p^\mu\frac{\partial F}{\partial x^\mu}-\Gamma^i_{\alpha\beta}p^\alpha p^\beta\frac{\partial F}{\partial p^i}=0,
\end{equation}
where $F$ is defined on the future mass shell and $i=1,2,3$. This correspondence applies to coarse-grained observables: interference can remain in the instantaneous field density.

The approximation requires the local de Broglie wavelength $\lambdabar_{\rm dB}\equiv \hbar/|\boldsymbol{p}|$ to be much smaller than the scales over which the macroscopic distribution and geometry vary~\cite{Widrow:1993qq, Widrow:1996eq}. For a Schwarzschild MBH, stable bound orbits have~$\mathcal L\geq\sqrt{12}\,R_M$, where $\mathcal L$ is the specific angular momentum. Their local spatial momentum (as measured by static observers) obeys $|\boldsymbol p|\geq\mb\mathcal L/r$, and therefore,
\begin{equation}
    \frac{\lambdabar_{\rm dB}}{r}\leq\frac{\hbar}{\mb\mathcal L}\leq\frac{\lambdabar_C}{\sqrt{12}\,R_M}.
\end{equation}
%
%The curvature length scales as $(r^3/R_M)^{1/2}$ and is of order $r$ or larger in the region occupied by these orbits. 
Thus, $\lambdabar_C\ll R_M$ gives a sufficient condition for a coarse-grained collisionless description of the spike.

In the adiabatic formation scenario, MBH growth contracts a pre-existing halo while approximately conserving the actions of regular bound orbits. The conservation of the relevant actions preserves their semiclassical character $(\gg\hbar)$, extending the effective particle description back to the initial halo orbits. Therefore, the ultralight boson distribution can be evolved using the same collisionless adiabatic construction of Ref.~\cite{Sadeghian:2013laa}.

%
%%%%%%%%%%%%%%%%%%%%%%%
\subsection{Particles behaving as waves}
%%%%%%%%%%%%%%%%%%%%%%%
%

%
\begin{figure*}
    \centering
    \includegraphics[width=\linewidth]{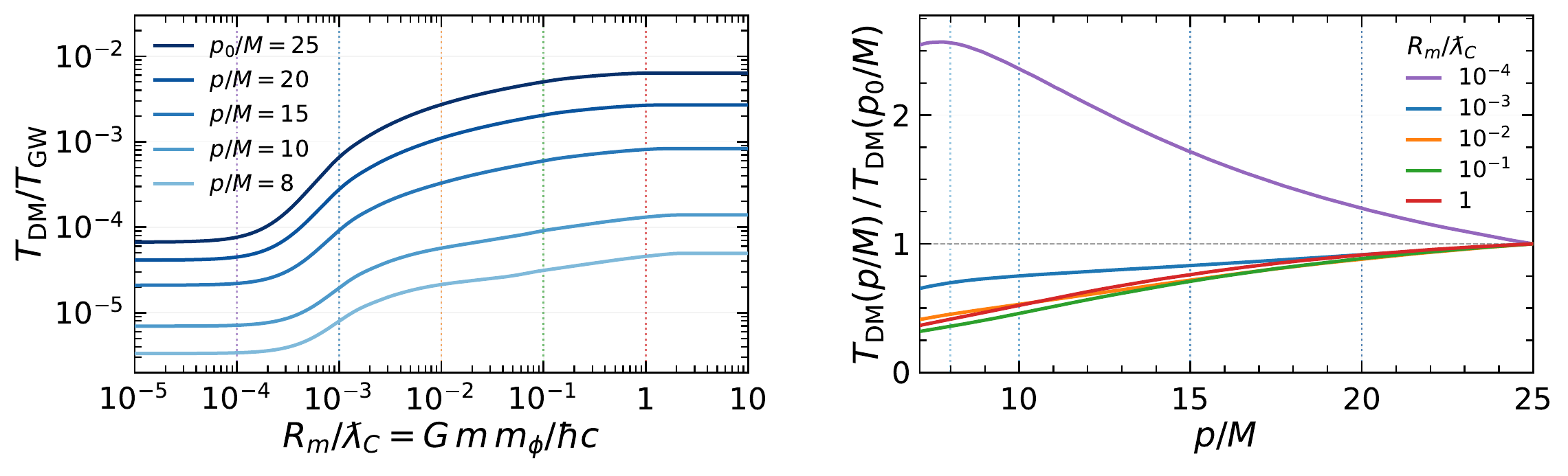}
    \caption{\textbf{Environmental torque from a dark matter spike.} \textit{Left}: Ratio of the dark matter to the gravitational-wave torque, as a function of $R_m/\lambdabar_C$, for the fiducial EMRI ($M=2\times10^5\,M_\odot$, $m=18\,M_\odot$) and a spike normalisation $\bar\rho=1$. The ratio scales linearly with $\bar\rho$. Curves correspond to orbital separations ranging from $p_0=25\,M$, near the start of the observation, to $p=8\,M$, near plunge. The torque depends on the DM particle mass only through $R_m/\lambdabar_C$ in the range $10^{-4}\lesssim R_m/\lambdabar_C \lesssim 1$. Dotted lines mark the values injected in Fig.~\ref{fig:mass-posteriors}. \textit{Right}: Radial profiles of the dark matter torque, normalised to their value at $p_0=25\, M$, for the same particle masses (matching colours). Dotted lines mark the orbital separations shown in the left panel.}
\label{fig:torque-ratio}
\end{figure*}

Although ultralight bosons can behave as collisionless particles on the scale of the primary, their interaction with the companion can remain sensitive to their wavelength. We describe this interaction locally as scattering by a Schwarzschild BH of mass $m$, with the incident velocity $v^{\hat i}$ measured in the companion's rest frame. We treat the incident distribution as locally homogeneous and regulate the long-range gravitational interaction using the Hill-scale prescription $b_{\max}=r q^{1/3}$, with $r$ the companion's orbital radius. The relevant dimensionless parameters are
\begin{equation}
    \alpha\equiv\frac{R_m}{\lambdabar_C}, \qquad
    \beta\equiv\frac{b_{\rm gr}}{\lambdabar_{\rm dB}}=\alpha \gamma_v\frac{1+v^2}{v},
\end{equation}
where $b_{\rm gr}=m(1+v^2)/v^2$ is the characteristic gravitational deflection impact parameter, and $\gamma_v\equiv(1-v^2)^{-\nicefrac{1}{2}}$.

We distinguish absorption by the companion from momentum transfer through gravitational deflection (i.e. dynamical friction). In the low-frequency limit, $\alpha \gamma_v\ll1$, the absorption cross section is~\cite{Vicente:2022ivh, Traykova:2023qyv}
\begin{equation}
    \sigma_{\rm acc}^{\rm w}
    =\frac{16\pi m^2}{v}
    \frac{2\pi\beta}{1-e^{-2\pi\beta}}.
\end{equation}
For $\beta\ll1$, the effective momentum transfer cross section associated with deflection takes the form~\cite{Hui:2016ltb, Vicente:2022ivh, Traykova:2023qyv}
\begin{equation}
    \sigma_{\rm scat}^{\rm w}
    =4\pi b_{\rm gr}^2
    \left[\operatorname{Cin}(X)+\frac{\sin X}{X}-1\right],
    \quad X\equiv\frac{b_{\max}}{\lambdabar_{\rm dB}},
\end{equation}
where $\operatorname{Cin}(X)=\int_0^X(1-\cos t)\,\mathrm{d}t/t$. 

In the collisionless particle limit, absorption and deflection are determined by timelike Schwarzschild geodesics and become independent of $\mb$, as follows from the weak equivalence principle. We denote the corresponding cross sections by $\sigma_{\rm acc}^{\rm p}=\pi b_{\rm cr}^2(v)$ and $\sigma_{\rm scat}^{\rm p}=4\pi b_{\rm gr}^2\Lambda_{\rm p}(v,b_{\max})$, where $b_{\rm cr}$ separates captured from scattered trajectories and $\Lambda_{\rm p}$ includes relativistic deflection effects and is given explicitly in Ref.~\cite{Vicente:2025gsg}. 

To connect the wave and particle regimes, we use
\begin{equation}
    \sigma_A(v,\alpha)
    =\min\!\left[\sigma_A^{\rm w}(v,\alpha),
    \sigma_A^{\rm p}(v)\right],
    \quad A\in\{\mathrm{acc},\mathrm{scat}\},
\end{equation}
with the dependence on $b_{\max}$ understood. This prescription extrapolates the asymptotic wave expressions beyond their strict domains of validity and caps each contribution at its particle value. This phenomenological prescription is motivated by the numerical results of Ref.~\cite{Traykova:2023qyv}.

In the companion's local frame, the four-acceleration is obtained by integrating the incident momentum flux weighted by these cross sections over the local spike velocity distribution perceived by the companion (as in~\cite{Vicente:2025gsg}),
\begin{equation}\label{eq:accelera}
	{a}^{\hat \mu} = \frac{1}{m}\int \frac{\dd N}{\dd V \dd^3 v} v \left[\sigma_{\rm acc}(v,\alpha) + \sigma_{\rm scat}(v,\alpha)\right]p^{\hat \mu}_\perp \dd^3 v\,,
\end{equation}
with $p^{\hat \mu}_\perp\equiv\mb\gamma_v(0,v^{\hat i})$. For quasi-circular EMRIs around a Schwarzschild MBH, the specific torque is written in terms of the four-acceleration as in \cite[Eq.~8]{Vicente:2025gsg}.

Figure~\ref{fig:torque-ratio} shows how the environmental torque depends on the DM particle mass, parametrised by $\alpha$, and the semi-latus rectum $p$ (equal to the orbital radius for circular orbits). The torque weakens for lighter DM particles (left panel), as wave effects suppress dynamical friction~\cite{Hui:2016ltb}. At $\alpha\lesssim10^{-4}$, however, it reaches a plateau dominated by accretion drag, which becomes independent of the particle mass in the low-mass limit~\cite{Unruh:1976fm}. Although our prescription enforces only continuity across the transition between the wave and particle regimes, the resulting torque is nearly smooth, consistent with the numerical results of Ref.~\cite{Traykova:2023qyv}.
The DM particle mass also changes the radial profile of the environmental torque (right panel), allowing it to be constrained independently of the spike normalisation.

%%%%%%%%%%%%%%%%%%%%%%%
\subsection{Universality of the inner radial profiles}
%%%%%%%%%%%%%%%%%%%%%%%

The environmental torque entering our inspiral model depends on the phase-space distribution of the DM spike and so, in principle, on the properties of the initial halo. Here, we show that this dependence can be approximately captured by a single normalisation over the radial ranges relevant to our analysis ($r\lesssim30\, R_M$). Both the density and torque profiles (for a fixed DM particle mass) have nearly universal radial shapes.

The adiabatic growth of a central MBH approximately preserves the actions of regular bound orbits and the distribution function expressed in those actions~\cite{Gondolo:1999ef,Sadeghian:2013laa,Ferrer:2017xwm}. For a specified initial distribution and final MBH mass, the surviving orbits are thus insensitive to the detailed growth history in the adiabatic regime. We first derive the scaling with halo normalisation and MBH mass, and then assess numerically the residual dependence of the density and torque profiles on the initial cusp slope.

We describe a stationary, phase-mixed, spherically symmetric distribution around a Schwarzschild MBH by
\begin{equation}
    F(x^\mu,p_\nu)= \mb^{-3}f(\mathcal{E},\mathcal{L})\, ,
\end{equation}
where $\mathcal E=-p_t/\mb$ and $\mathcal L$ are the conserved energy and total angular momentum per unit particle rest mass. The invariant distribution $F$ is normalised by
\begin{equation}
    \dd N=F(-p_\mu n^\mu)\,\dd V\,\dd P,
\end{equation}
where $n^\mu$ is the future-directed unit normal to the spatial volume element $\dd V$. In a local orthonormal frame, $\dd P=\dd^3p/p^{\hat 0}$ on the future mass shell $p^\mu p_\mu=-\mb^2$.

We model the initial halo within a reference radius $r_s$ as an isotropic, spherical cusp,
\begin{equation}
    \rho_i(r)=\rho_s\left(\frac{r}{r_s}\right)^{-\gamma}\, , \qquad 0<\gamma<2\, ,
    \label{eq:initial-cusp}
\end{equation}
with $\rho_s\equiv\rho_i(r_s)$. We assume that this cusp dominates the initial gravitational force over the orbits that supply the inner spike. Its enclosed mass is
\begin{equation}
    M_i(<r)=M_s\left(\frac{r}{r_s}\right)^{3-\gamma}, \quad M_s=\frac{4\pi\rho_s r_s^3}{3-\gamma},
\end{equation}
and the relative potential is
\begin{equation}
    \Psi(r)=\Psi_s+c_s\left[1-\left(\frac{r}{r_s}\right)^{2-\gamma}\right], \quad c_s\equiv\frac{M_s}{(2-\gamma)r_s},
    \label{eq:cusp-potential}
\end{equation}
where $\Psi_s\equiv\Psi(r_s)$. A spherical continuation outside $r_s$ changes the additive constant $\Psi_s$, but not the force inside the cusp. We adopt $r_s=100\,{\rm pc}$ as a reference radius.

For an isotropic halo, the initial distribution function $f_i(E')$, normalized by $\rho_i=\mb\int f_i\,\dd^3v$, follows from Eddington inversion \cite{binney},
\begin{equation}
    f_i(E')=\frac{1}{\sqrt{8}\,\pi^2 \mb} \int_0^{E'}\frac{\dd\Psi}{\sqrt{E'-\Psi}} \frac{\dd^2\rho_i}{\dd\Psi^2}\, ,
    \label{eq:eddington-cusp}
\end{equation}
with $E'\equiv\Psi-v^2/2$.
The density and torque in the inner spike are dominated by particles whose initial orbits lie well within $r_s$. Their distribution can then be evaluated using the inner-cusp asymptotic form. Defining $\delta\equiv1-(E'-\Psi_s)/c_s$, we obtain, as $\delta\to0$,
\begin{equation}
    f_i(E')\simeq \frac{\rho_s} {\mb \,(2\pi c_s)^{3/2}} \frac{\Gamma\!\left(\frac{6-\gamma}{2(2-\gamma)}\right)}{\Gamma\!\left(\frac{\gamma}{2-\gamma}\right)}\, \delta^{-\frac{6-\gamma}{2(2-\gamma)}}\,,
    \label{eq:inner-cusp-df}
\end{equation}
The unspecified outer halo enters only through terms that are subleading as~$\delta\to0$.
We then grow a Schwarzschild MBH of mass $M$ adiabatically at the centre of the DM cusp. Conservation of phase-space density gives
\begin{equation}
    f_{\rm sp}(\mathcal E,\mathcal L) = f_i\!\left[E'_*(\mathcal E,\mathcal L)\right]\,,
    \label{eq:spike-df-map}
\end{equation}
where $E'_*(\mathcal E,\mathcal L)$ is obtained from the conservation of the radial action as described in Ref.~\cite{Sadeghian:2013laa}, and $f_{\rm sp}$ is set to zero for plunging orbits.

The scaling of the spike distribution function with $\rho_s$ and $M$ can be obtained analytically. We first note that the radial action per unit particle rest mass, $\mathcal{J}_r$, of the particles in the initial cusp can be expressed as
\begin{equation}
    \frac{\mathcal J_r}{r_s \sqrt{c_s}\, \delta^{\frac{4-\gamma}{2(2-\gamma)}}} =  j\bigg(\frac{\mathcal{L}}{r_s \sqrt{c_s}\, \delta^{\frac{4-\gamma}{2(2-\gamma)}}};\gamma\bigg)\,.
\end{equation}
This implies that the initial energy expressed as function of the orbital actions is homogeneous, 
\begin{equation}
    \delta(\lambda\, \mathcal{J}_r, \lambda\, \mathcal L)=\lambda^{\frac{2(2-\gamma)}{4-\gamma}} \delta( \mathcal{J}_r, \mathcal L)\,.
\end{equation}
Therefore, expressing $\mathcal J_r/M$ as function of $\mathcal{E}$ and $\mathcal{L}/M$, we can write $\delta$ as
\begin{equation}
    \delta=\Big(\frac{M}{r_s \sqrt{c_s}} \Big)^{\frac{2(2-\gamma)}{4-\gamma}} \Delta_\gamma\Big(\mathcal{E}, \frac{\mathcal L}{M}\Big)\,.
\end{equation}
Combining these relations gives
\begin{equation}
    \mb M^2 f_{\rm sp}(\mathcal E,\mathcal L)
    \simeq
    (\rho_s r_s^\gamma)^{\frac{1}{4-\gamma}}
    M^{\frac{2-\gamma}{4-\gamma}}
    H_\gamma\!\left(\mathcal E,\frac{\mathcal L}{M}\right),
\end{equation}
where the dimensionless function $H_\gamma\propto \Delta_\gamma^{-\frac{6-\gamma}{2(2-\gamma)}}$ contains the remaining dependence on the cusp slope. 

The nontrivial dependence of $H_\gamma$ on the initial cusp slope could, in principle, propagate to the radial profiles of the density and torque. However, when performing the relevant integrals in velocity-space [e.g. Eq.~\eqref{eq:accelera}] we find that, due to loss-cone effects in the spike distribution function, both quantities have an approximately universal shape over the radii relevant to our analysis ($r\lesssim 30\,R_M$). The dependence on the initial cusp can be approximately captured by a simple normalisation. That allows us to get both quantities from their reference profiles. For the environmental torque
\begin{equation}
    \dot l_{z,\mathrm{env}}(r;q,\alpha)
    \simeq
    \bar\rho\,\dot l_{z,\mathrm{env}}^{\mathrm{ref}}[(M_0/M) r; q, \alpha],
\end{equation}
where the reference torque is evaluated for the reference spike and MBH discussed in the text. 

The normalisation is well described by
\begin{equation}
    \bar\rho\simeq\gamma\, g(\gamma)\left(\frac{\rho_{100\,{\rm pc}}}{150\,\mathrm{GeV}/c^2\,\mathrm{cm}^{-3}}\right)^{\frac{1}{4-\gamma}}\left(\frac{M}{10^6\,M_\odot}\right)^{\frac{2-\gamma}{4-\gamma}}.
    \label{eq:DensityNorm}
\end{equation}
Over the tested range \(0.25\leq\gamma\leq1.5\), the remaining slope dependence is fitted by $g(\gamma)\simeq(1+2.4\gamma^2+0.8\gamma^6)/4.2$; for this range of slopes, the maximum fractional residual in the torque within $r\leq 30 R_M$ is less than $5\%$. 

%%%%%%%%%%%%%%%%%%%%%%%
\subsection{Waveform model and data analysis setup}
%%%%%%%%%%%%%%%%%%%%%%%

\paragraph*{Waveform model.} For the EMRI waveforms, we use the FastEMRIWaveforms (\texttt{few}) package~\cite{Chua:2020stf, Katz:2021yft, Speri:2023jte}, a modular, GPU-accelerated framework that generates EMRI waveforms in hundreds of milliseconds. Its latest release implements a fully relativistic adiabatic model for eccentric, equatorial EMRIs in a Kerr background~\cite{Chapman-Bird:2025xtd}. We restrict to quasi-circular, equatorial orbits around non-spinning MBHs. The modularity of \texttt{few} lets us add the torque induced by the dark matter spike directly to the evolution of the companion's specific angular momentum
\begin{equation}
    \dot l_z=\dot l_{z,{\rm GW}}(r)+\bar\rho\,\dot l^{\,\rm ref}_{z,\rm env}\big[(M_0/M)\,r;q,\alpha\big]\,,
\end{equation}
which is linear in the normalisation $\bar\rho$, with $\alpha\equiv m\,\mb/\hbar=R_m/\lambdabar_C$. We evolve the inspiral under this modified flux balance and generate the waveform along the resulting trajectory. The torque model is too slow for on-the-fly evaluation during Bayesian parameter estimation, so we precompute it for the reference spike ($\bar\rho=1$), in units of the MBH mass and per unit mass ratio $q=m/M$, on a grid of $119$ separations in $7\le r/R_M\le30$, $19$ mass ratios in $10^{-5}\le q\le10^{-4}$ and $61$ values of $-5\le\log_{10}\alpha\le1$, and interpolate with tricubic splines during inference. To validate that the grid is sufficiently dense, we rebuilt the interpolant from a grid twice as coarse in $q$ and $\log_{10}\alpha$ and compared it with the omitted grid values. The relative difference in the torque, averaged over the radial range as in Ref.~\cite{Fantoccoli:2026idl}, has a median of $2\times10^{-4}$. The full grid is more accurate still, since the error of a cubic interpolant decreases steeply with the grid spacing $h$ (more precisely, as $\propto h^4$). For $\alpha\gtrsim2$ the tabulated torque no longer depends on $\alpha$ (particle limit). We set the environmental torque to zero for $r<7.1\,R_M$, just above the inner edge of the grid. The cutoff does not affect our results, since it typically concerns the last couple of days before plunge, when the gravitational-wave torque exceeds the environmental one by more than five orders of magnitude, compared to only 2 or 3 orders of magnitude at the beginning of observation (see Fig.~\ref{fig:torque-ratio}).

The vacuum inspiral depends on the MBH mass $M$, the companion mass $m$, the MBH spin parameter $a$ (here zero), the initial separation $p_0$ and the initial azimuthal phase $\Phi_0$. The signal also depends on two angle pairs, $(\theta_K,\phi_K)$ for the orientation of the orbital angular momentum and $(\theta_S,\phi_S)$ for the sky location in the Solar System barycentre frame, together with the luminosity distance $d_L$. We call $(M,m,p_0,\Phi_0)$ the \textit{intrinsic} parameters, $(\theta_K,\phi_K,\theta_S,\phi_S,d_L)$ the \textit{extrinsic} parameters, and $(\bar\rho,\alpha)$ the \textit{environmental} parameters. We consider systems observed up to plunge, which are the most likely to be detected because the final stage of the inspiral dominates the signal-to-noise ratio~\cite{Speri:2026ade}.

\paragraph*{Parameter estimation framework.} We did a fully Bayesian analysis to determine how well the boson mass and the spike density can be constrained simultaneously. Our goal is to obtain the posterior distribution for the EMRI parameters $\lambda$ given the data $d$, $p(\lambda\mid d)\propto\mathcal L(d\mid\lambda)\,\pi(\lambda)$, in terms of the likelihood $\mathcal L(d\mid\lambda)$ and the prior $\pi(\lambda)$. Since we are interested only in parameter estimation, the evidence acts as a normalisation constant and can be neglected. We adopt a standard Gaussian likelihood
\begin{align}
    \ln\mathcal L(d\mid\lambda)&=-\frac12\sum_{X=A,E}\braket{d_X-h_X(\lambda)\mid d_X-h_X(\lambda)}\,, \\
    \braket{a\mid b}&=4\,{\rm Re}\int_{f_{\min}}^{f_{\max}}\frac{\tilde a^*(f)\,\tilde b(f)}{S_n(f)}\,\dd f\,,
\end{align}
where $d$ is the injected signal, $h(\lambda)$ the waveform model, $\lambda$ the full parameter set (intrinsic, extrinsic and environmental), and $S_n(f)$ the one-sided power spectral density (PSD) of LISA. The detector response is computed with second-generation time-delay interferometry on realistic LISA orbits~\cite{Katz:2022yqe}, and we analyse the channels $A$ and $E$, treating their noise as independent. Signals are tapered with a Tukey window of parameter $0.005$ to reduce spectral leakage, and the PSD is the SciRDv1 sensitivity~\cite{Babak:2021mhe} including the Galactic confusion foreground. A particular noise realisation shifts the posterior maximum away from the injected values without appreciably broadening the distribution, corresponding to a loss of accuracy rather than precision. For this reason, we limit this study to noiseless realisations. Real LISA observations will exhibit non-Gaussian noise, instrumental glitches and data gaps, all of which complicate the likelihood and make its evaluation more costly~\cite{Burke:2025bun, Boumerdassi:2025gvf, Burke:2026rjq}.

\paragraph*{Sampler and priors.} We sample the posterior with the Markov-chain Monte Carlo sampler \texttt{eryn}~\cite{Karnesis:2023ras}, using $64$ walkers at each of two parallel-tempered rungs with fixed inverse temperatures $\beta=1$ and $\beta\simeq0.49$, for a faster exploration of parameter space. The swap acceptance between the two temperatures is $22$--$27\%$. We sample eleven parameters, the intrinsic $(\ln M,m,p_0,\Phi_0)$, the extrinsic $(\cos\theta_K,\phi_K,\cos\theta_S,\phi_S,d_L)$ and the two environmental ones, where the masses are in detector frame and $p_0$ is the separation at the start of the observation.

The correlation between the two environmental parameters, discussed in the main text, turns the posterior into a long, curved ridge along which the likelihood is strongly non-Gaussian (as also found in Ref.~\cite{Fantoccoli:2026idl} for accretion disks). We therefore sample in coordinates that follow the ridge. For each injection, we first compute the curve of best-fitting $\bar\rho$, $M$, $m$, $p_0$ and $\Phi_0$ as a function of $\alpha$ with the orbital phase-based profile described below, and then sample $\log_{10}\alpha$ together with the offsets of $\log_{10}\bar\rho$, $\ln M$, $m$, $p_0$ and $\Phi_0$ from that curve. 

The priors used are the following: for the extrinsic and intrinsic parameters uniform in the sampled coordinates within five standard deviations determined by a Fisher matrix approximation (computed with the \texttt{StableEMRIFisher} package), truncated to their physical ranges, which we verified a posteriori to contain the posterior in all case; over $-5\le\log_{10}\alpha\le1$, the range of the torque grid; and over $[-5,2]$ for the offset of $\log_{10}\bar\rho$, which covers all range of the posteriors obtained. The walkers start within $0.03$ Fisher standard deviations of the injection in all coordinates except $\log_{10}\alpha$, which is spread uniformly along the ridge over a range covering the expected posterior. Proposals are drawn from two affine-invariant stretch moves~\cite{Goodman:2010dyf}: one acting on all eleven parameters, with probability $0.7$, and one acting on $\log_{10}\alpha$ alone, which moves the walkers along the ridge, with probability $0.3$. Their acceptance fractions are $15$--$32\%$ and $44$--$69\%$, respectively, across all cases studied.

\begin{figure*}
    \centering
    \includegraphics[width=\linewidth]{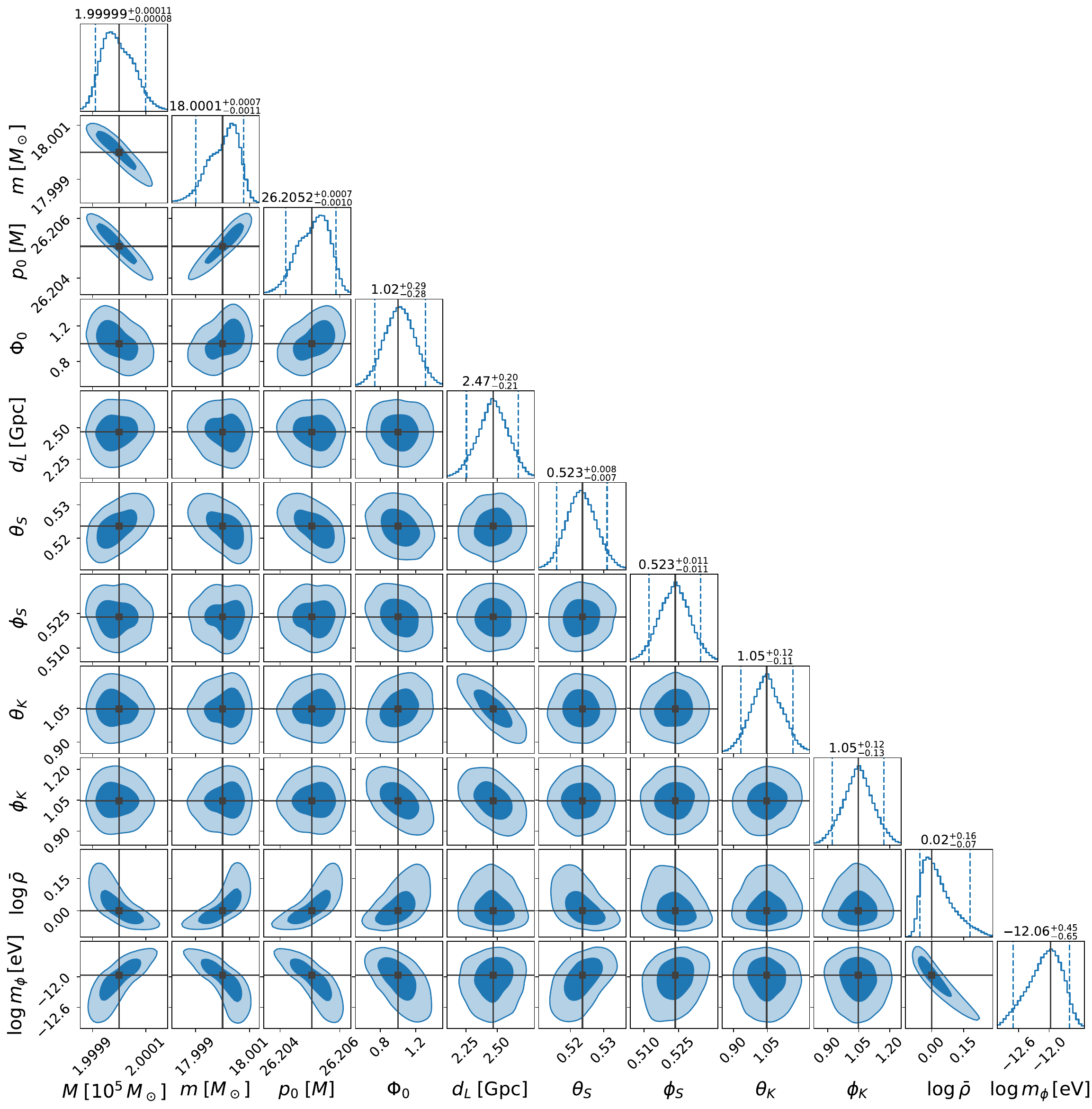}
    \caption{\textbf{Full posterior for the fiducial EMRI with \boldmath$m_\phi=1.1\times10^{-12}\,{\rm eV}$.} 1D and 2D marginal posteriors of the intrinsic, extrinsic and environmental parameters for the corresponding injection of Fig.~\ref{fig:mass-posteriors}. Contours enclose $50\%$ and $90\%$ of the posterior. Dashed lines mark the $90\%$ credible intervals, also quoted with the medians above each panel, and solid lines the injected values. The extrinsic parameters are close to Gaussian, whereas the posterior is non-Gaussian along the $\bar\rho$--$m_\phi$ degeneracy.}
    \label{fig:full-posterior}
\end{figure*}

After burn-in, each posterior chain spans about $>50$ integrated autocorrelation times, which we adopt as a convergence criterion, and we also verify that the two halves of each chain give
consistent posteriors. As an example, we show in Fig.~\ref{fig:full-posterior} the marginalised 2D posteriors across all eleven EMRI parameters for the fiducial system explored in Fig.~\ref{fig:mass-posteriors} and a boson mass  $m_\phi=1.1\times10^{-12}\,{\rm eV}$.

\paragraph*{Dephasing-based estimates.} To map the constraints of Fig.~\ref{fig:measurable-band} without full MCMC runs, we estimate the likelihood from the orbital phase alone. For noiseless data, $d=h_0\equiv h(\lambda_0)$, the Gaussian likelihood above has its maximum at the injected values $\lambda_0$, and a template $h\equiv h(\lambda)$ has
\begin{align}
    \Delta\ln\mathcal{L}&=-\frac12\braket{h-h_0\mid h-h_0}\nonumber\\
    &=\braket{h\mid h_0}-\frac12\Big(\braket{h\mid h}+\braket{h_0\mid h_0}\Big)\,.
\end{align}
The environmental torque changes the inspiral slowly, so a template with a different boson mass has almost the same amplitude evolution as the injected signal, while the small change in the orbital evolution accumulates in its phase over the many cycles in band. Then $\braket{h\mid h}\simeq\braket{h_0\mid h_0}={\rm SNR}^2$, and $\Delta\ln\mathcal{L}\simeq-\big({\rm SNR}^2-\braket{h\mid h_0}\big)$ is $-{\rm SNR}^2$ times the mismatch between the two signals. For quasi-circular orbits, the signal is dominated by the quadrupole harmonic, which carries more than $90\%$ of ${\rm SNR}^2$ for the systems we consider (averaged over sky orientations), so we write $h_0\simeq A(t)\cos2\Phi(t)$ and $h\simeq A(t)\cos[2\Phi(t)+2\Delta\Phi(t)]$, where $\Phi$ is the orbital phase of the injection and $\Delta\Phi(t)$ the difference between the orbital phases of the template and of the injection. Since the frequency evolves slowly, each frequency is emitted at essentially a single instant (the so-called stationary-phase approximation), and 
\begin{equation}
    \braket{h\mid h_0}\simeq\int\frac{A^2(t)}{S_n\big(f(t)\big)}\cos\big(2\Delta\Phi(t)\big)\,\dd t\,,
\end{equation}
and the ${\rm SNR}^2$ is the same integral with $\Delta\Phi=0$. Defining $w(t)\equiv A^2/[S_n(f)\,{\rm SNR}^2]$, the fraction of the squared signal-to-noise ratio accumulated per unit time ($\int w\,\dd t=1$), we obtain
\begin{align}
    \Delta\ln\mathcal{L}&\simeq-{\rm SNR}^2\Big[1-\int w(t)\cos\big(2\Delta\Phi(t)\big)\,\dd t\Big]\\
    &\simeq-2\,{\rm SNR}^2\!\int w\,\Delta\Phi^2\,\dd t\,,
\end{align}
where the second form holds for small dephasing. We evaluate $w$ with the leading-order amplitude $A\propto\dot\Phi^{2/3}$ and neglect the slow modulation from the detector response, so that the extrinsic parameters enter only through the SNR. 

For each injected boson mass, we therefore integrate the inspiral in a spike with $\bar\rho=1$ and, for each displaced value of $\alpha$, minimise $\int w\,\Delta\Phi^2\,\dd t$ over $\bar\rho$, $M$, $m$, $p_0$ and $\Phi_0$, with an iterative least-squares fit, started from the best fit at the previous value of $\alpha$ so that the fit follows the $\bar\rho$--$m_\phi$ degeneracy. The posterior on $\log_{10}\alpha$ follows from $\exp(\Delta\ln\mathcal{L})$ and the flat prior on $\log_{10}\alpha\in[-5,1]$, and we quote the width of its $90\%$ credible interval. 

The MCMC analyses, which include the higher harmonics and the full detector response, test these approximations together. For the fiducial system, the estimated widths agree with those of the posteriors of Fig.~\ref{fig:mass-posteriors} to within $20\%$ for all five injections. We also ran full MCMC analyses for four further configurations at ${\rm SNR}=50$, varying the masses, observation time and spike density: $10^5+7\,M_\odot$ over $3$ years with $\bar\rho=1$, $2\times10^5+18\,M_\odot$ over $3$ years with $\bar\rho=3$, $2\times10^5+14\,M_\odot$ over $4$ years with $\bar\rho=3$, and $10^5+4\,M_\odot$ over $3$ years with $\bar\rho=0.3$. In all four, the estimate follows the $\bar\rho$--$m_\phi$ degeneracy found by the MCMC, and the widths of the $90\%$ intervals agree to within $5\%$ in three cases, while for $2\times10^5+18\,M_\odot$ with $\bar\rho=3$ the MCMC interval is $30\%$ wider. The dephasing-based estimate therefore reproduces the order-of-magnitude of the width of the credible interval at a small fraction of its computational cost, which makes it a practical tool to assess the measurability of $m_\phi$ with EMRIs across parameter space.

\subsection{Acknowledgments}

RV, TK and GB gratefully acknowledge the support of the Dutch Research Council (NWO) through an Open Competition Domain Science-M grant, project number OCENW.M.21.375. The work of FD and GB was supported by the European Research Council (ERC) Advanced Grant \textit{De Tenebris} (grant agreement No. 101198798), funded by the European Union. Views and opinions expressed are however those of the author(s) only and do not necessarily reflect those of the European Union or the European Research Council Executive Agency. Neither the European Union nor the granting authority can be held responsible for them.

% \newpage
\bibliography{refs}

\end{document}